# A universal framework to efficiently share material and process resources in the DNA construction world

Hideto Mori[1–3,5] and Nozomu Yachie[1-5,]*

[1]Research Center for Advanced Science and Technology, The University of Tokyo, Tokyo 153-8904, Japan
[2]Institute for Advanced Biosciences, Keio University, Tsuruoka 997-0035, Japan
[3]Graduate School of Media and Governance, Keio University, Fujisawa, Kanagawa 252-0882, Japan
[4]School of Biomedical Engineering, Faculty of Applied Science and Faculty of Medicine, The University of British Columbia, Vancouver, BC, Canada
[5]Twitter: @yachielab

*Correspondence should be addressed to N.Y. (nozomu.yachie@ubc.ca).

**DNA constructs and their annotated sequence maps have been rapidly accumulating with the advancement of DNA cloning, synthesis, and assembly methods. Such a resource has the potential to be optimally utilized in an autonomous DNA building platform. However, most DNA design processes today remain manually operated with the assistance of graphical user interface (GUI) software. Furthermore, as seen commonly in the life sciences, reproducibility of DNA construction process descriptions is usually not guaranteed, and utilization of previously developed materials and protocols is not appropriately credited. Here, we developed an open-source process description and resource sharing framework QUEEN (a framework to generate quinable and efficiently editable nucleotide sequence resources) to resolve these issues in building DNA. QUEEN enables the flexible design of new DNA by using existing DNA resource files and recoding the construction process in an output file (GenBank file format). The GenBank files generated by QUEEN are able to regenerate the process codes that perfectly clone themselves and bequeath the design history to successive DNA constructs that recycle their partial resources. QUEEN-generated GenBank files are compatible with the existing DNA repository services and software. We propose QUEEN as a solution to start significantly advancing our material and protocol sharing of DNA resources.**

Designing and building DNA are necessary processes in most biological research today. The introduction of exogenous DNA into cells and animals allows for monitoring of molecular and cellular behaviors, and reverse engineering and functional enhancement of target systems. Chemical DNA synthesis[1] and assembly methods[2] have largely advanced in the last couple of decades, leading to the whole synthesis of bacterial[3,4] and yeast chromosomes[5-7] and the establishment of biofoundries towards the automated production of engineered microorganisms[8,9]. Today, a total of multiple hundred thousand DNA plasmids have been deposited to public DNA repository services, such as AddGene[10] and DNASU[11]. While the growing DNA resources have been successfully accelerating biological research, current DNA resource sharing and building methods have major room for improvements in optimal recycling and crediting of DNA resources and reproducible protocols.

Previously established DNA resources have yet to be fully utilized to produce new DNA constructs efficiently. Some "DNA brick" systems have been proposed with a limited set of compatible restriction digestions and ligations as a platform to optimize the DNA construction process[12-15]. However, they have not been widely used by the community, probably because the number of compatible parts remains to meet the diverse demands in biology, or because PCR-based fragment preparation and highly specific overlap DNA assembly methods[16-21] have become the mainstream and freed molecular cloning from enzymatic assembly methods. We instead believe that the key to advancing the reusability of the rapidly accumulating DNA resources is not the standardization of DNA module assemblies, but an advanced process design system, whereby the most optimal construction process of a target DNA is autonomously formulated with a combination of maximal use of existing DNA resources and chemical DNA synthesis. We are, however, one step away from starting to establish such a system because there is no standardized process description system for building DNA. Most of us have still been designing DNA manually with GUI software tools, and the protocols have been described in natural language.

Additionally, as commonly seen in biology, no efficient system to share reproducible protocols has been developed for building DNA. While this issue will be partially resolved by switching from natural language protocols to a standardized process description system defining restrictions and requirements to describe DNA building, it needs to be coupled with another system that validates its reproducibility, and a strong motivation is required to transfer the biological research community to such a new system. Furthermore, as also commonly seen in biology but especially for DNA materials whose partial modules and construction knowledge have been inherited from study to study, we have yet to establish a mechanism to credit resource and protocol developers[8,9]. Software development and version control systems like GitHub would achieve a similar system in sharing DNA materials and their construction protocols; however, such a system in building DNA has not been established. Lastly, no universal toolkit to describe and simulate a range of changes in DNA sequences with their annotations has been developed. While the methods to build and program DNA have diversified, with CRISPR genome editing and base editing technologies providing new ways of efficiently editing DNA sequences[22], current DNA editing software tools cannot assist users to incorporate these new methods or allow them to develop new plugins.

A quine in computer science is a program which replicates a copy of itself without a need of any inputs. The intriguing concept of quine that first appeared in the mid-19th century provides us fruitful thoughts on self-replicating machines and their potentials. In this study, inspired by this concept, we developed a simple versatile framework QUEEN (a framework to generate "quinable" and efficiently editable nucleotide sequence resources) that has a potential to resolve all the above-mentioned issues in describing, sharing, and crediting DNA building processes and DNA resources. QUEEN is a Python package where DNA building processes and dynamic DNA sequence changes can be freely described and simulated. Products of QUEEN-generated DNA products can be output in the common GenBank (gbk) DNA annotation file format. Due to the flexibility of the gbk file format, the QUEEN-generated gbk files achieve several unique features. First, the QUEEN-generated gbk files can regenerate the quine codes that generated themselves using QUEEN (Fig. 1a). Just like the research community has already been doing, DNA materials can be deposited to a DNA repository service together with their gbk files created by QUEEN (Fig. 1b). While it serves as a regular gbk file, the file's construction process information can also be recovered by QUEEN, where the reproducibility of the protocol is certified by this clonability. Second, in designing a new DNA construct using QUEEN, when DNA parts are taken from existing QUEEN-generated gbk files (Fig. 1c) or a protocol generated from a QUEEN-generated gbk file is modified (Fig. 1d), the newly produced gbk file inherits the source material information as well as their annotations and previous production process histories, allowing for a progressive tracking of the recycling of materials and methods. This feature will be a backbone for establishing new ways of evaluating genetic resources, protocols, and developers based on how they are inherited in the community (Fig. 1e). QUEEN-generated gbk files can be spread throughout the community as they can be treated as a regular gbk file by existing software tools. This feature will prime the promotion of the new material and protocol sharing model. Furthermore, the issues we try to solve here have many commonalities with those in other domains of biological experiments. The framework we propose here for DNA resources could trigger a discussion on establishing similar frameworks outside of DNA construction experiments, or, ideally, a unified framework for the entire natural science with a generalized process semantic.

**QUEEN**
QUEEN is implemented as an open-source Python module (Fig. 1f). It was designed so that software tools and databases will be developed to enhance its basic capabilities and the momentum of support for the idea behind this framework in the biological research community. In QUEEN, the QUEEN class is provided to represent double-stranded (ds)DNA objects with their annotated sequence features and editing histories. Each sequence feature is given by the "DNAfeature" class. A new QUEEN class object can be generated by specifying a DNA sequence or importing a sequence file in gbk or FASTA file format. A circular DNA sequence like a bacterial DNA plasmid can also be defined as a QUEEN class object, where a circular permutation system maintains nucleotide positions of the dsDNA. A QUEEN class object generated or edited by a QUEEN program can be output as a gbk file that encodes the quine code to replicate the DNA manipulation process.

In QUEEN, dsDNA objects can be manipulated by four basic operational functions: "cutdna", "modifyends", "flipdna", and "joindna", that can collectively represent any of the standard molecular DNA cloning processes, two search functions, "searchsequence" and "searchfeature", and two super functions, "editsequence" and "editfeature". A linear or circular dsDNA can be segmented into multiple fragments or linearized by cutdna. Each cut site can be defined by a single DNA sequence position (blunt-end cut), two sequence positions for both strands (sticky-end cut), or a DNA feature object holding a "cut_site" qualifier that can represent any restriction enzyme site. Any DNA feature objects such as gene annotations are inherited to the corresponding regions of the newly generated fragments. DNA feature objects on the cut boundaries are also passed to the new fragments, each with a qualifier specifying it as a broken feature. For convenience, we also implemented "cropdna" to extract a segment sandwiched by two cut sites. Any number of dsDNA objects can be assembled at once by joindna, which requires the connecting DNA end structures to be compatible (i.e., only blunt ends and compatible sticky ends can be joined). If the assembly reconstitutes any sequences that previously contained any other DNA feature objects in the generating dsDNA, those objects are also autonomously restored. Any DNA sequence can be added to or removed from whichever strand of DNA ends by the modifyends operation. While this operation serves as an alternative way to create restriction cut site-compatible DNA ends, it also enables the description of any overlap-based DNA assembly methods requiring long overlapping sequences on DNA ends by making the overlapping sequences sticky ends. dsDNA fragments can also be flipped by the flipdna operation. This operation can be used to model not only the flipping of dsDNA fragments for their assemblies with others, but also DNA sequence inversions via site-specific recombination systems, such as Cre-loxP and FLP-FRT.

When a new DNA is constructed manually, the user often does not know or remember the complete information of all the DNA parts being treated. searchsequence searches for a queried sequence provided by regular expression in the target dsDNA and returns DNA feature objects of the matched sequence region. It allows fuzzy matching with user-defined options. PCR can be modeled by searching two primer sequences in a template DNA and generating a dsDNA fragment cropped by outer positions of the primer binding sites. Restriction enzyme digestion can be simulated by searching for DNA sequences matching the DNA feature objects describing the target restriction cut patterns in a target DNA and using the same DNA feature objects to digest the target DNA. DNA feature object is composed of an attribute-information structure for their "feature_id", "feature_type", and user-defined "qualifiers." searchfeature can be operated against any of the attribute types or their corresponding DNA sequences by querying a search term by regular expression and specifying fuzzy matching options. Similar to searchsequence and searchfeature, editsequence and editfeature can search for target sequences and DNA features, but they can also enable direct editing of the QUEEN dsDNA objects. editsequence enables direct replacement of sequences matched to a query with a specified sequence. This function can be employed to model genome editing. editfeature enables creation of a new DNA feature and its attributes, and the removal and editing of existing features. Same as when a DNA feature object is fragmented by cutdna, the DNA feature objects corresponding to a DNA sequence region modified by editsequence receive notification flags. If the DNAfeature objects are not valid for the new sequence, the user can modify them using the editfeature function.

To export and display information of QUEEN dsDNA objects, QUEEN provides five output functions: "outputgbk", "printsequence", "printfeature", "visualizemap", and "visualizeflow". Newly constructed QUEEN dsDNA objects can be exported into gbk files by outputgbk. printsequence is a standard output function that displays a dsDNA sequence of a given QUEEN object with modified end structures. printfeature outputs a well-formatted table for user-selected attribute values in DNAfeature objects in a QUEEN object. visualizemap and visualizeflow generate an annotated sequence map and an operational history flow chart of a given QUEEN object, respectively, whose visualization styles can be flexibly modified by the user. These output functions support users in easily programming a QUEEN code with an interactive programming environment, such as Jupyter Notebook, where the users can promptly check the progress of a DNA construction (links to examples provided in Methods).

QUEEN progressively records all operations provided by the four basic operational functions, two search functions, and two edit functions, into a history attribute of a DNA feature object defined for the entire QUEEN class object. This history attribute stores not only the operational history for the present DNA construction, but also past operational histories of the parental QUEEN objects whose partial fragments are inherited to the present construction. Therefore, while a QUEEN-generated gbk file can be operated by other non-QUEEN supported tools, it also has abilities to produce a quine code that self-replicates and provides the information of how previous DNA materials have been manipulated and inherited to the present DNA construct. Furthermore, users can group a subset of operational flows and provide a narrative description of methodological procedures in the quinable script. This function allows DNA builders to provide experimental procedures in natural language along with their process semantic descriptions with minimal loss of information. It also enables the generation of the "Materials and Methods" description for a DNA construct from its QUEEN-generated gbk file. (Note that the further development of GUI-based software tools would allow users to program the DNA construction process without making them aware of the QUEEN programming semantic.)

**Simple molecular cloning**
We demonstrated that QUEEN enabled the description of various DNA construction processes, simulation of dynamic DNA programs, and production of quinable annotated DNA files. We first tested a description of a simple gene cloning procedure to derive a lentiviral plasmid pRS112 that we constructed previously[23] (Fig. 2a). An enhanced green fluorescent protein (eGFP)-encoding cassette was first amplified from the pLV-eGFP plasmid by PCR with primers that have overhang sequences encoding EcoRI and BamHI restriction digestion sites. The amplified PCR product and the destination plasmid pLV-SIN-CMV-Puro were both digested by EcoRI and BamHI and ligated to obtain the final product.

This entire process could be described using QUEEN with eleven operational steps (Fig. 2b). To obtain the PCR product, the 18-bp 3' sequences of the two primers were each subjected to searchsequence to obtain their annealing sites. The internal DNA sequence flanked by the primer annealing sites was obtained by cropdna, followed by tethering of the primer sequence to both ends by modifyends, representing the final PCR product. EcoRI and BamHI cut sites and digestion patterns were defined and searched by searchsequence in both the PCR product and destination lentiviral plasmid. The DNA feature objects defining the cut sites were used for the double digestion by cutdna. Finally, the digested fragments with compatible sticky ends were connected by joindna. We confirmed that the generated sequence was identical to that of the previously generated gbk file, and the QUEEN-generated gbk file could produce a quine code for the same plasmid.

**Deep resource inheritance**
To demonstrate QUEEN can describe overlap DNA assembly and produce gbk files that can track their building histories and those of inherited DNA parts, we replicated the construction process of six CRISPR base-editor plasmids using QUEEN: pCMV-Target-AID, pCMV-Target-ACE, pCMV-AIDmax, pCMV-Target-ACEmax, pCMV-BE4max(C), and pCMV-ACBEmax that we constructed previously[23] (Fig. 3a). The entire construction processes of these plasmids relied on PCR amplification of DNA fragments and Gibson Assembly, where, upon some plasmid constructions, their DNA parts were recycled for other plasmid constructions. In brief, PCR fragment preparations were described using searchsequence, cropdna, and modifyends, as described above. Gibson Assembly reactions were described by generating long compatible sticky ends using modifyends and assembling them by joindna.

Two fragments and one fragment were amplified from pcDNA3.1_pCMV-nCas-PmCDA1-ugi pH1-gRNA(HPRT) and pCMV-ABE7.10, respectively, and assembled to generate pCMV-Target-AID. Different fragments, one each from these templates, were amplified and used to generate pCMV-Target-ACE. After synthesizing human codon-optimized PmCDA1-UGI cloned into pUC-optimized-PmCDA1-ugi, single fragments were amplified from each of pCMV-BE4max, pCMV-ABEmax, and pUC-optimized-PmCDA1-ugi, and assembled to derive pCMV-Target-AIDmax. The same fragment from pCMV-ABEmax and another fragment from the same plasmid were also assembled with the codon-optimized PmCDA1-UGI fragment from pUC-optimized-PmCDA1-ugi to derive pCMV-Target-ACEmax (Fig. 3b). Finally, three new fragments were amplified from pCMV-BE4max and assembled with a part of the assembled fragment of pCMV-Target-AIDmax to obtain pCMV-BE4max(C), and with a part of the assembled fragment of pCMV-Target-ACEmax to obtain pCMV-ACBEmax. Throughout these assemblies, we sometimes amplified adjacent fragments separately and assembled them back in the same order in a destination plasmid for better PCR amplification of higher-quality, shorter fragments rather than facing difficulties in amplifying longer PCR products. We demonstrated that even if a DNA region having DNA feature objects was once truncated by this operation, QUEEN successfully restored the features in the assembled DNA when an assembly reconstituted the original sequence (Fig. 3c).

We confirmed that the constructed QUEEN scripts could generate target DNA sequences identical to the previously constructed ones. Their quine codes and complete operational histories could also be produced from the generated gbk files (Supplementary Figs. 1 and 2). Notably, pCMV-BE4max(C) and pCMV-ACBEmax inherited DNA fragments from pCMV-Target-AIDmax and pCMV-Target-ACEmax, respectively, by importing their gbk files. We observed in the process histories of the secondary plasmids that the construction processes of the first plasmids were entirely inherited (Supplementary Fig. 2). Furthermore, the construction processes of pCMV-Target-ACEmax and pCMV-ACBEmax were similar to those of pCMV-Target-AIDmax and pCMV-BE4max(C), respectively, where two and three fragments were shared. Therefore, we also generated another two gbk files for pCMV-Target-ACEmax and pCMV-ACBEmax by retrieving the quine codes from the pCMV-Target-AIDmax and pCMV-BE4max(C) gbk files, respectively, and directly editing them. The gbk files generated through this editing strategy successfully recorded the partial inheritance of the previous protocols (Supplementary Fig. 3), showing that QUEEN can create a platform to track the partial and full inheritance of DNA materials and construction protocols.

**Simulation of dynamic DNA changes**
As seen in yeast mating-type cassette switching, site-specific DNA recombination, meiotic chromosomal recombination, and genome editing, DNA is not a static object but can act dynamically in a programmed manner. Although highly efficient genome editing methods can be used to construct DNA plasmids[24, 25] and synthetic genetic circuits[26, 27], both involving DNA sequence changes, no standard framework is proposed to simulate progressing changes in annotated dsDNA sequences. To demonstrate QUEEN can be a platform to simulate and share such dynamic DNA circuits, we implemented the genetic six-input, one-output Boolean Logic Look Up Table (LUT) of BLADE[28] into QUEEN (Fig. 4a). In this circuit, upon input of site-specific DNA recombinases, the circuit DNA sequence alters by multi-step deletions and/or inversions of segments sandwiched by corresponding recombinase target sequences. The input patterns of four recombinases (Vica, B3, PhiC31, and Bxb1) configures one of the 16 Boolean logic gates where the remaining two recombinases (Cre and FLP) serve as two input signals to the logic gate (Fig. 4b). The output is given as GFP expression derived by the promoter located upstream of the circuit DNA. The intertwined segment recombinations yield one or none of the four GFP genes to be expressed, depending on the signal input pattern.

We implemented DNA segment deletion by cutdna and joindna and inversion by cutdna, flipdna and joindna. After obtaining a gbk file describing the initial state of the DNA circuit, we loaded the DNA object in a QUEEN script and simulated its behaviors for all of the 64 possible signal input patterns. All the input patterns conferred the expected DNA sequence outcomes by annotating the active GFP gene with other expected feature annotations (Fig. 4c). We also output the resulting DNA sequences to a gbk file and demonstrated that its quine code and process history could be derived from the output file (Supplementary Fig. 4).

**Discussion**
There have been several software packages developed to design molecular cloning processes and generate annotated plasmid files, such as Ape (https://jorgensen.biology.utah.edu/wayned/ape/), Benchling (https://www.benchling.com/), Geneious[29], j5[30], Pydna[31], Raven[32], and SnapGene (https://www.snapgene.com/). While most of them are GUI-based software tools, including Benchling, a cloud-based web application service, Pydna is a Python programming package that enables the description of DNA construction processes. Notably, the operational functions of these tools are all implemented for specific cloning methods, such as Gibson Assembly, Golden Gate Assembly, and traditional enzymatic digestion and ligation cloning. Some of them[30, 32] even have functions to suggest several of the best cloning options for target DNA construction process. These specified functions seem intuitive and convenient for users as long as they design DNA materials with the assumed methods but lack the elasticity to incorporate new DNA building methods. In contrast, we hypothesized that DNA cloning operations can be generalized by the combination of four basic operations: "cutdna", "joindna". "modifyends", and "flipdna". We demonstrated that this system could describe the equivalent operations of the previous tools in various examples. QUEEN is not limited to designing the plasmid construction process and production of their annotated DNA files. The gbk file format is common to store genomic sequences. While a range of organisms whose genomic sequences are available are genetically engineered with targeted gene deletion and transgene insertion through homologous DNA repair, Cre-loxP, and genome editing, no software tool to describe these processes has been developed. Although whole-cell simulation[33, 34] is still far from simulating all cellular metabolic reactions purely from the genomic sequence at the single-nucleotide resolution, a platform to describe and edit the annotated genomes and their evolution is indispensable towards such a goal. The simulation of the BLADE circuit demonstrated that QUEEN is capable of flexibly simulating a dynamic process of DNA sequence alternations. There have only been tools that allows users to design and simulate gene expression patterns of synthetic genetic circuits with gene regulatory networks[35-37] and a limited set of genetic alternations[38]. In combination with the existing simulation platforms, QUEEN could also accelerate the construction of genetic circuits and engineered cells in systems and synthetic biology fields.

Recording operational history in the output gbk file and inheritance of such by the descendant gbk files are also unique features of QUEEN that have great potential to change the ways of building DNA. The ability to generate a quine code from a QUEEN-generated gbk file enables the accompaniment of protocols with the material data describing the products, and thus certifies their reproducibility. A protocol retrieved from a QUEEN-generated gbk file can be edited to generate a new DNA construct. This process can also be recorded and passed to a producing gbk file. Benchling and SnapGene enable recording of simulated DNA construction processes and allows different users to access such information. However, these software features are limited only within the software environments. They are unable to share the process histories widely with the research community and track the deep inheritance of materials and protocols from one DNA construct to another. Before QUEEN, no practical system to pair and share DNA material data with their reproducible protocols and parental DNA material and protocol resources was proposed. If the research community starts generating gbk files using QUEEN-compliant software tools and DNA repository services, and begins maintaining databases for these files with web search application programming interfaces (APIs), there will be new ways of evaluating research and researchers beyond citation count or H-index. The impacts of a material and its protocol and developer can be evaluated not only by a "trending" notification or how many times it is requested by other researchers, like shown in Addgene, but also by how it is inherited in successive products even over multiple generations. Furthermore, once such DNA resources have sufficiently accumulated, it would enable the development of an AI-assisted DNA design system; a building process of user-requested DNA could be autonomously designed with the most optimal recycling of previous products and with the best context-dependent DNA cloning strategies suggested on the previously described processes; the recycling strategy could take into account their availabilities in the user's laboratory and delivery logistics from DNA repository services; the cloning methods could also be selected based on how many times the community has succeeded in similar methods. These ideas could be implemented to gene synthesis and assembly automation systems[39]. The current implementation of QUEEN requires output gbk files to inherit all of the parental resource information, making their file sizes large. This implementation is unnecessary as such information can be stored separately in a cloud server and linked from the newly generating gbk files. However, we believe this is currently the best form so that QUEEN-generated resources can easily penetrate the existing DNA repository services and the community, before the community realizes their value and are encouraged to develop tools and databases for this new framework. (We do not believe that designing an extensive system that is only possible by establishing all necessary subsystems is the right way to go, but rather, establishing tricks that users can accept independently and gradually with the smallest possible hurdles is important to achieve such a vision.)

A process ontology enables description of protocols with a certain resolution defined by the semantic system itself. A programming language, whose definition is narrower, enables a reproducible description of processes for their output products. QUEEN is a simple programming framework that certifies high-level descriptions of how DNA sequences are manipulated and the reproducibilities thereof. It does not allow a standardized description of experimental procedures from other aspects with instruments, apparatuses, reagents with their lots, and how they are manipulated with their environmental information, such as temperature, humidity, and gas consumption. This limitation is considered practically sufficient in DNA construction because the design and build processes are always DNA sequence-centric and the molecular cloning experiments are usually highly standardized with stable and reactive enzymes and well-established kits. Thus, the current implementation of QUEEN would achieve a large improvement in sharing DNA materials and DNA construction processes.

While it is important to establish a similar system for different experimental domains that require more complex descriptions on materials and processes, this is challenging. Ideally, laboratory automation of all procedures in natural science experiments could accompany the full semantic description of reproducible protocols[40] and alleviate many of the current issues, including the reproducibility crisis. However, significant technical and social contrivances would be needed primarily because the development of robotic systems requires tremendous investment and laboratory automation communities have not been progressing in concert (different projects develop their own systems which cannot easily be integrated). In QUEEN, one of the contrivances is the use of gbk file format so that the existing platforms can host them, allowing the biological community to smoothly start accumulating the new resources without any additional cost. For a versatile laboratory automation, a potential direction could be the development of a standardized "multiscale" process semantic, where any experimental processes of any resolution can be programmed— from the resolution of current material and methods description with researchers' best efforts to that for robotic executions with their APIs. It may be possible to promote the community to first use such a semantic for daily experiments with the support of a GUI-assisted editor, whose description can be shared, reused, and elaborated later for laboratory automation with robotics. This, if widespread, could accelerate the momentum of laboratory automation towards a similar vision that we propose here with QUEEN. Here, we described QUEEN as a tool to accelerate DNA construction, and as a concept that should be investigated across various experimental domains both within and outside the life sciences.

**Acknowledgments**
We thank the members of the Yachie laboratory at the University of British Columbia and the University of Tokyo for useful comments and discussions, especially Samuel King for proofreading and providing feedback on the manuscript. This study was performed under the Canada Research Chair program supported by the Canadian Institutes for Health Research (CIHR). H.M. was supported by the JSPS DC2 Fellowship and TTCK fellowship.

**Author contributions**
H.M. and N.Y. conceived the study and designed QUEEN. H.M. implemented QUEEN. H.M. and N.Y. wrote the manuscript together.

**Competing interests**
None declared.

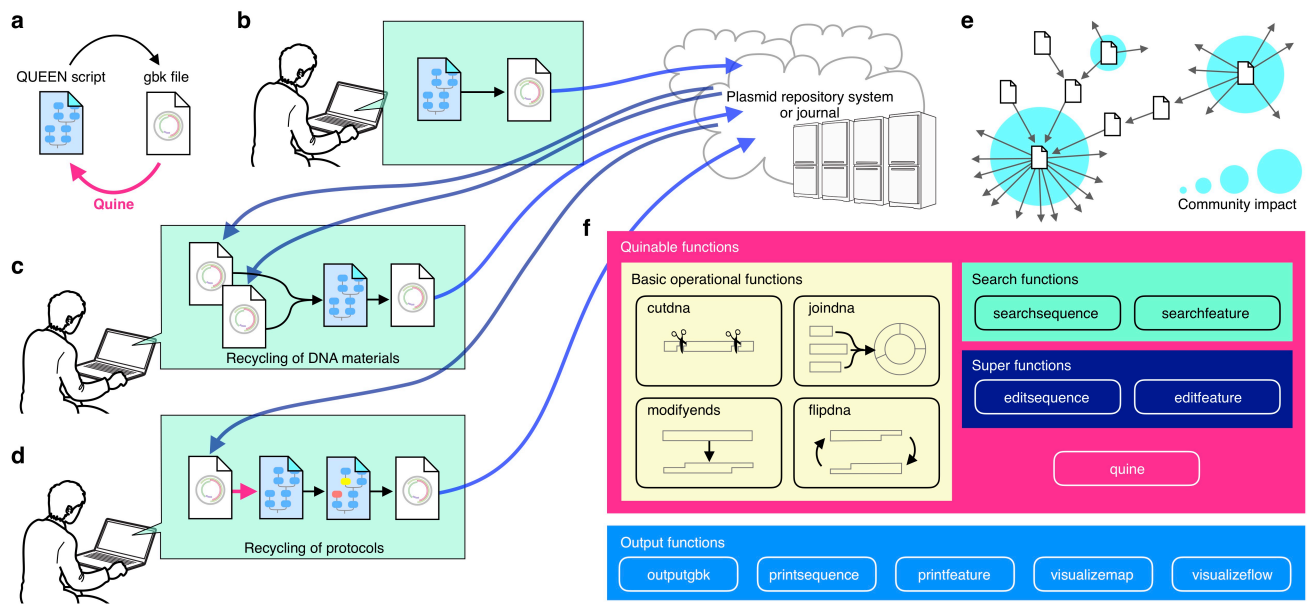

**Figure 1**
**QUEEN. a**, QUEEN enables the generation of a quine code from a gbk file. **b**, A QUEEN-generated gbk file can be deposited to an existing DNA repository as a regular gbk file. **c**, When DNA sequences are partially or fully inherited from previous QUEEN-generated gbk files to build a new DNA, the producing gbk file can contain all the production histories of the ancestral gbk files. **d**, When a QUEEN script obtained from a previously established gbk file is modified to design the building process of a new DNA, such a history is also inherited by the produced gbk file. **e**, A QUEEN-generated gbk file traces its previously utilized DNA materials and protocols, enabling the assessment of community impact of each DNA resource or developer. **f**, Overview of QUEEN operations.

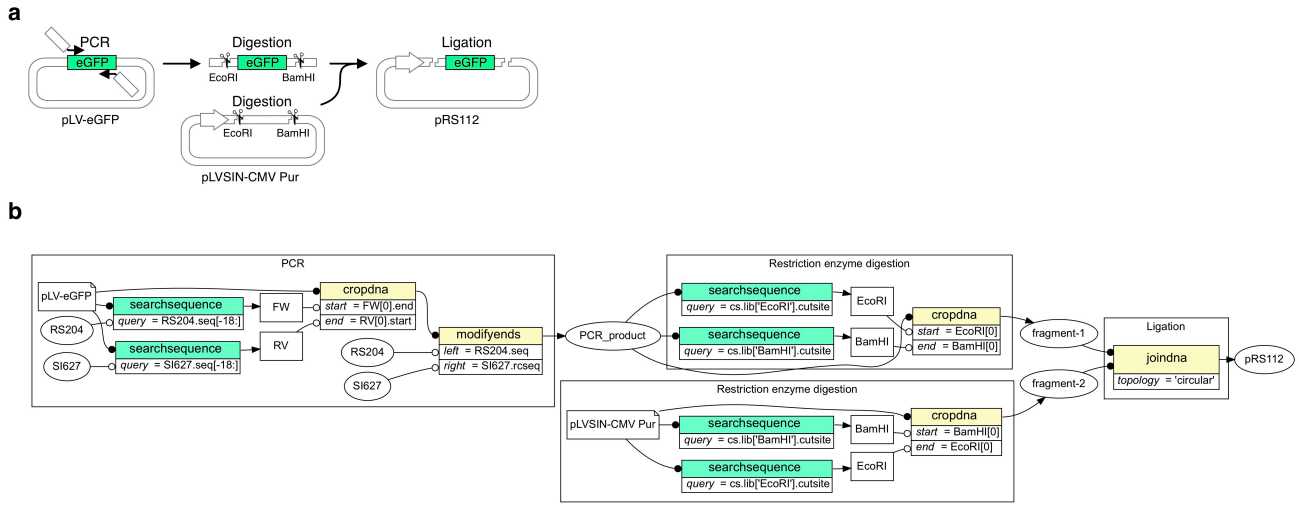

**Figure 2**
**Simple molecular cloning. a**, Conceptual diagram representing the cloning of eGFP into a lentiviral plasmid backbone. **b**, QUEEN operational flow representing the same DNA cloning process. The visualization was directly generated from the output gbk file by the visualizeflow function. The file shape, round, and uncolored rectangular objects represent the input gbk files, QUEEN objects, and DNAfeature objects, respectively. Colored boxes represent QUEEN operational functions with the colors corresponding to Fig. 1f. Open and closed circle-headed lines represent information flows as QUEEN objects and input parameters, respectively.

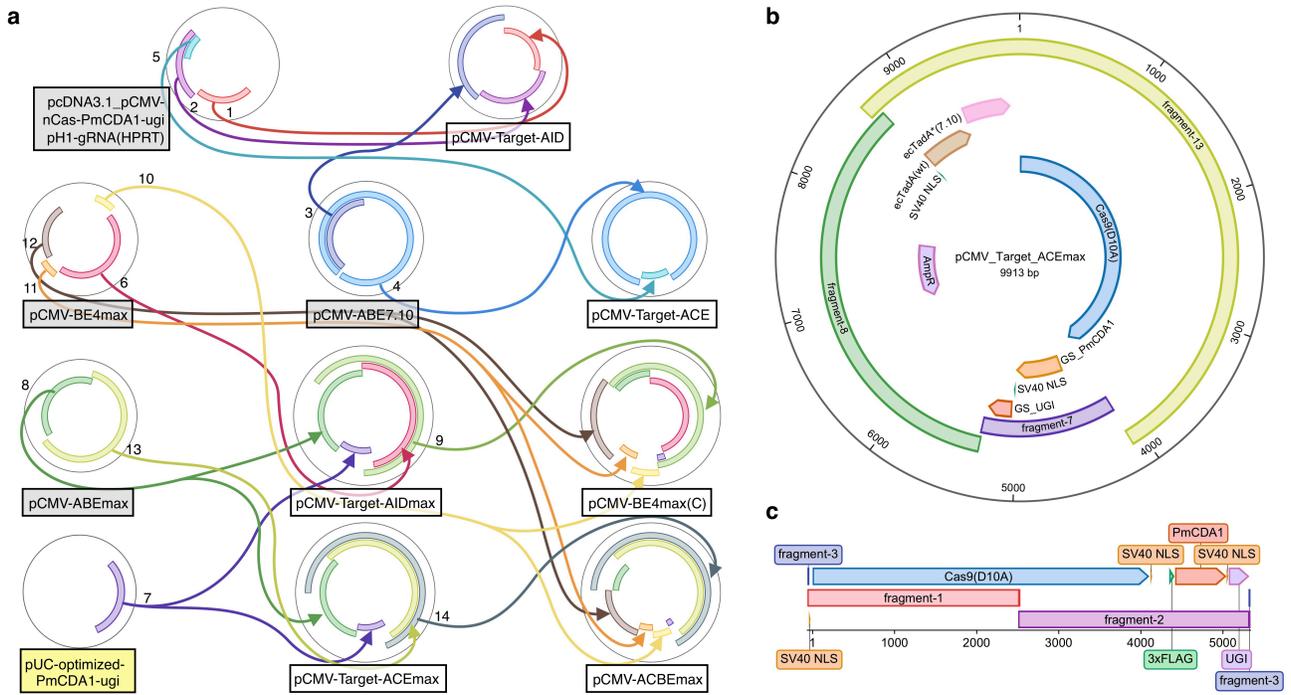

**Figure 3**
**Base editor plasmid constructions. a**, The construction lineages of base editor plasmids with recycling of their DNA sequences. The plasmid names with gray, yellow, and white boxes represent those obtained from Addgene, newly synthesized, or reconstructed using QUEEN, respectively. **b**, Annotated circular sequence map of pCMV-Target-ACEmax generated from the output gbk file by the "visualizemap" function. **c**, Recovery of "Cas9(D10A)" DNAfeature after joining two fragments each from a different DNA object having the same DNAfeature.

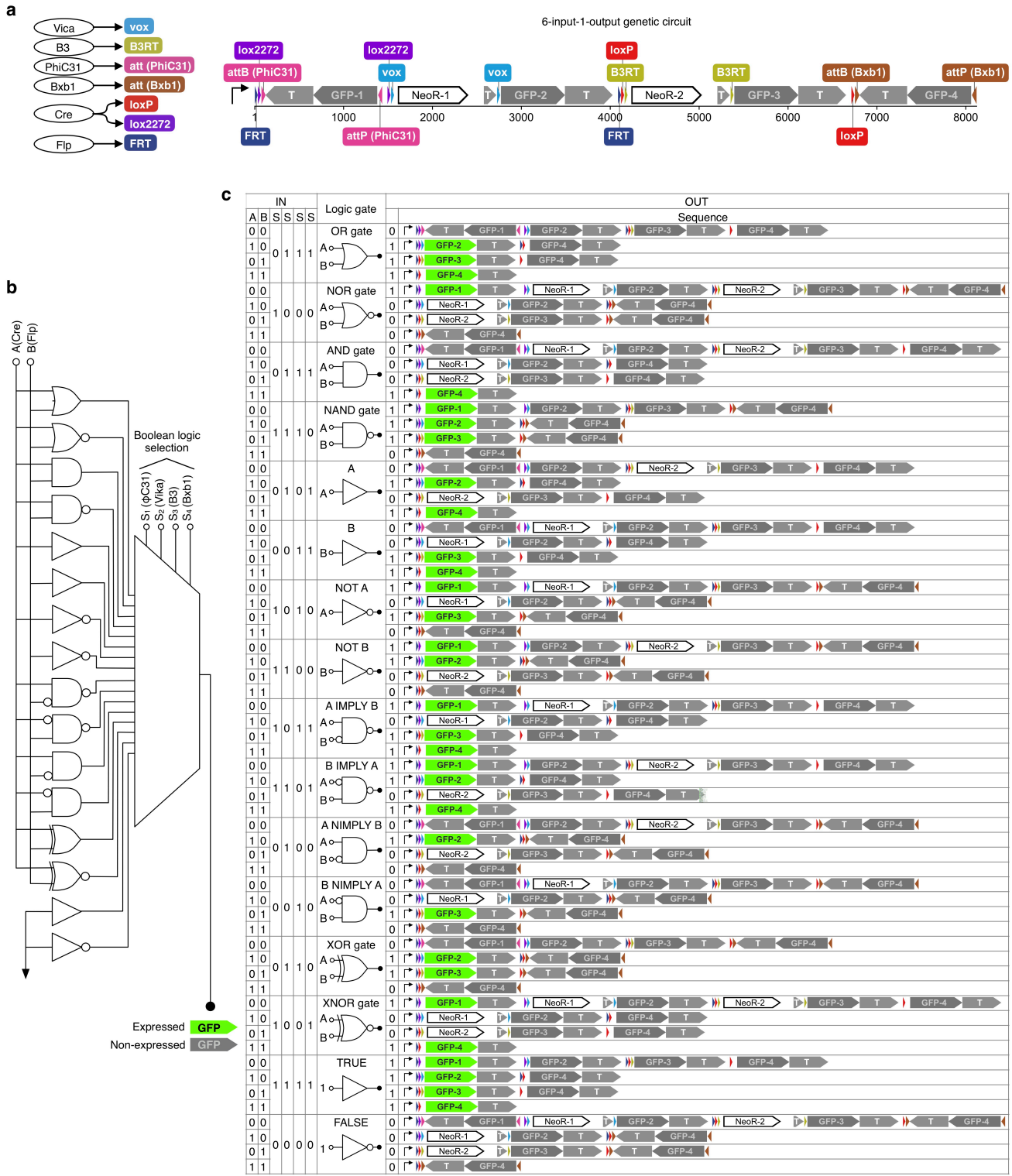

**Figure 4**
**Simulation of a Boolean logic LUT**. **a**, The six-input, one-output Boolean logic LUT of the BLADE system. Left, site-specific DNA recombinases (round nodes) and their corresponding target DNA sites (round rectangles). Right, the genetic Boolean logic LUT circuit. **b**, The wiring diagram of the genetic circuit. **c**,

Annotated DNA sequence outcomes of each signal input pattern simulated by QUEEN. GFP expresses if one of the GFP-encoding genes is placed in the same direction as the upstream promoter without being prevented by a transcription terminator.

## Methods

**Implementation of QUEEN.** QUEEN was implemented as a Python 3.7 module. It requires BioPython[41] mainly for GenBank file parsing. The visualization of annotated DNA sequence maps (visualizemap) and operational process flow charts (visualizeflow) from QUEEN objects are dependent on Python Matplotlib [42] and Graphviz[43] modules, respectively. The detailed usage of each function implemented in the QUEEN module is described in the GitHub repository (https://github.com/yachielab/QUEEN/blob/master/README.md). Example QUEEN scripts are provided as Jupyter Notebook files at https://github.com/yachielab/QUEEN/tree/master/demo/tutorial.

**Data availability.** Of the gbk files used as input files for the simulation of base editor plasmid constructions, pLV-eGFP, pCMV-ABE7.10, pcDNA3.1_pCMV-nCas-PmCDA1-ugi pH1-gRNA(HPRT), pCMV-BE4max and pCMV-ABEmax were obtained from Addgene (Plasmid IDs: 36083, 102919, 79620, 112903, and 112905, respectively). The gbk file for pLV-SIN-CMV-Puro was obtained from Takara Bio, Inc. (Japan). The gbk file for pRS112 and pUC-optimized-PmCDA1-ugi encoding the codon-optimized PmCDA1-UGI was created using Benchling. Some detail sequence feature annotations of input files were added manually before using them for the demonstration (the modified files are available at https://github.com/yachielab/QUEEN/tree/master/demo/sakata_et_al_2020). The gbk file used for the simulation of the Boolean logic LUT circuit was downloaded from Addgene (Plasmid ID 87554) to which sequence feature annotations for the site-specific recombination sites were added manually before the demonstration (the modified file is available at https://github.com/yachielab/QUEEN/tree/master/demo/Weinberg_et_aL_2017).

**Code availability.** QUEEN is an open-source software package distributed with GNU General Public License v3.0. The entire package and installation and user's manual are available at the GitHub repository (https://github.com/yachielab/QUEEN/). All of the source codes for QUEEN are placed in https://github.com/yachielab/QUEEN/tree/master/QUEEN. The QUEEN scripts used to construct the base editor plasmids and the simulation of the Boolean logic LUT are provided as Jupyter Notebook files at https://github.com/yachielab/QUEEN/tree/master/demo/sakata_et_al_2020 and https://github.com/yachielab/QUEEN/tree/master/demo/Weinberg_et_aL_2017, respectively.

All of the Jupyter Notebook files for the demonstrations in this study are available in GitHub and made executable in Google Colaboratory (Supplementary Table 1).

Supplementary Materials

# A universal framework to efficiently share material and process resources in the DNA construction world

Hideto Mori[1–3,5] and Nozomu Yachie[1-5,]*

[1]Research Center for Advanced Science and Technology, The University of Tokyo, Tokyo 153-8904, Japan
[2]Institute for Advanced Biosciences, Keio University, Tsuruoka 997-0035, Japan
[3]Graduate School of Media and Governance, Keio University, Fujisawa, Kanagawa 252-0882, Japan
[4]School of Biomedical Engineering, Faculty of Applied Science and Faculty of Medicine, The University of British Columbia, Vancouver, BC, Canada
[5]Twitter: @yachielab

*Correspondence should be addressed to N.Y. (nozomu.yachie@ubc.ca).

**Table of contents**

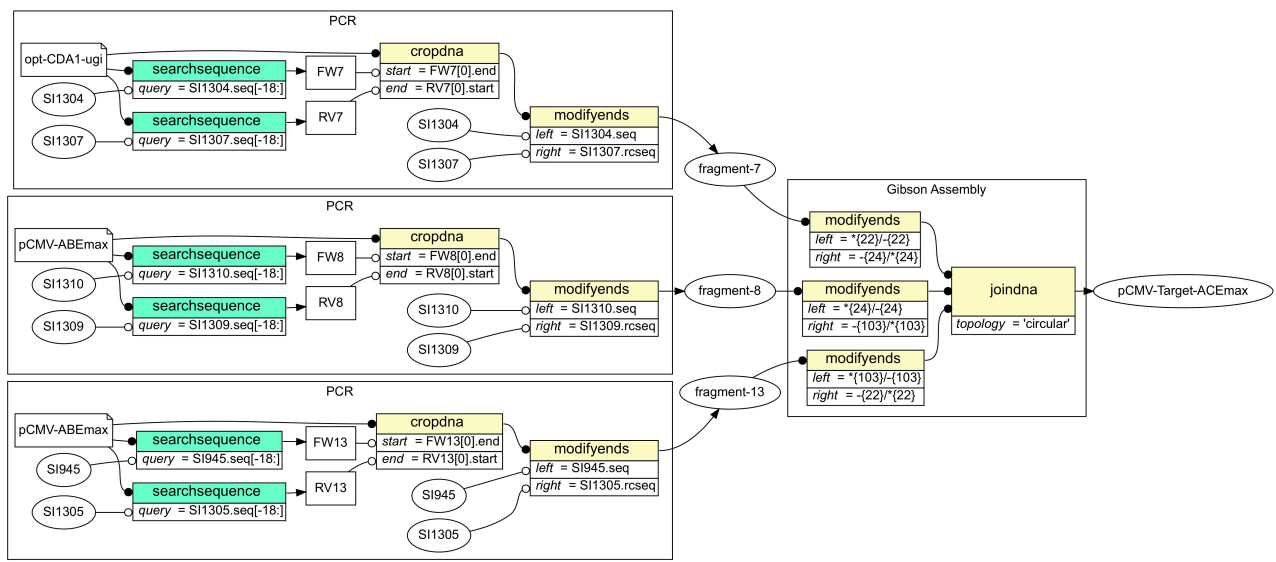

**Supplementary Figure 1**
**Operational process chart for pCMV-Target-ACEmax.** The visualizations were directly generated from the output gbk file simply by the visualizeflow function. The file shape, round and uncolored rectangular objects represent the input gbk files, QUEEN objects and DNAfeature objects, respectively. Colored boxes represent QUEEN operational functions with the colors corresponding to Fig. 1f. Open and closed circle-headed lines represent information flows as QUEEN objects and input parameters, respectively.

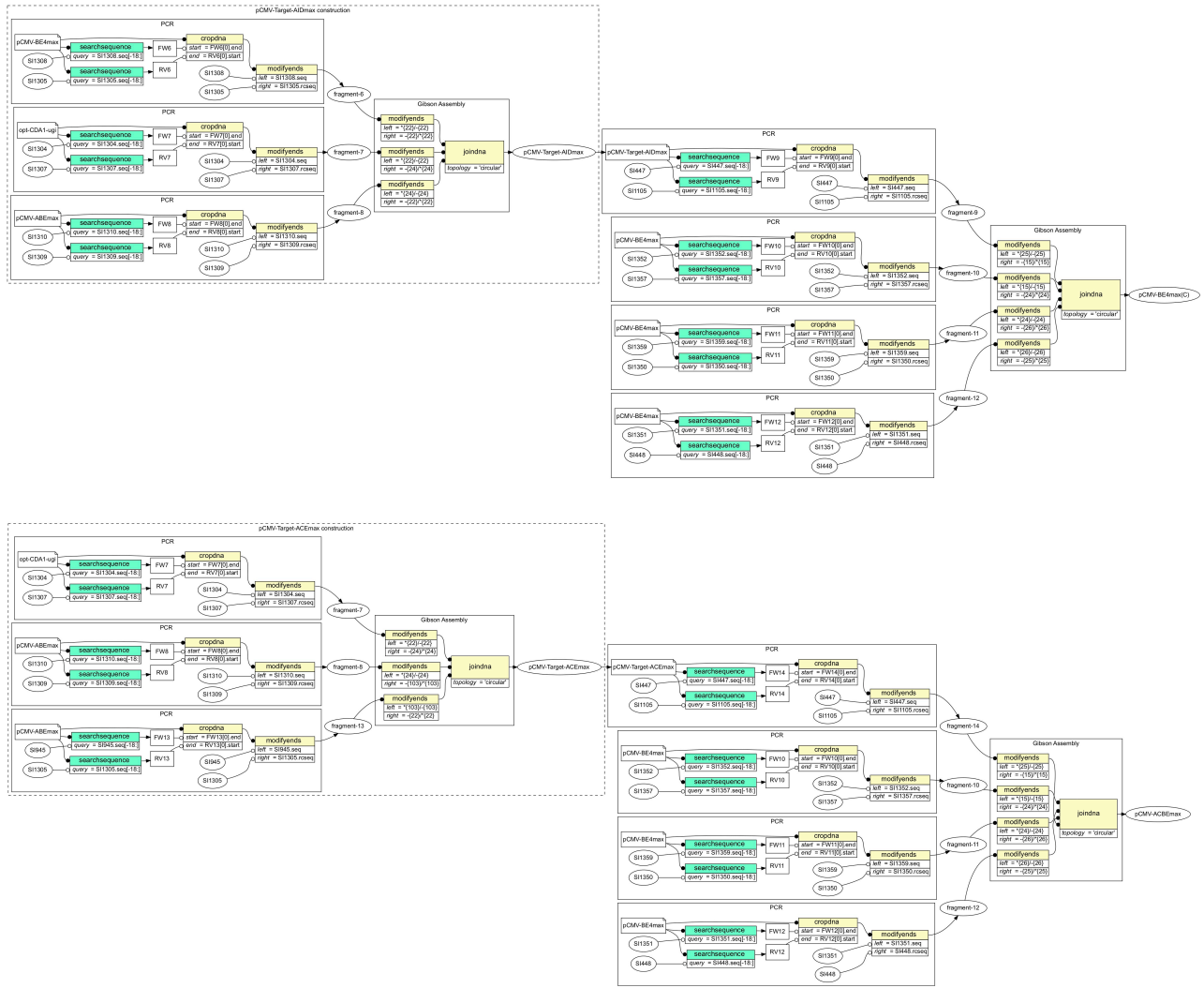

**Supplementary Figure 2**
**Operational process charts of QUEEN-generated plasmids derived by recycling previously established DNA objects.** The visualization was directly generated from the respective product gbk files by the visualizeflow function. See Supplementary Fig. 1 for detail. The operational histories enclosed by the dashed lines represent those of DNA objects previously generated by QUEEN and imported for the present DNA constructions.

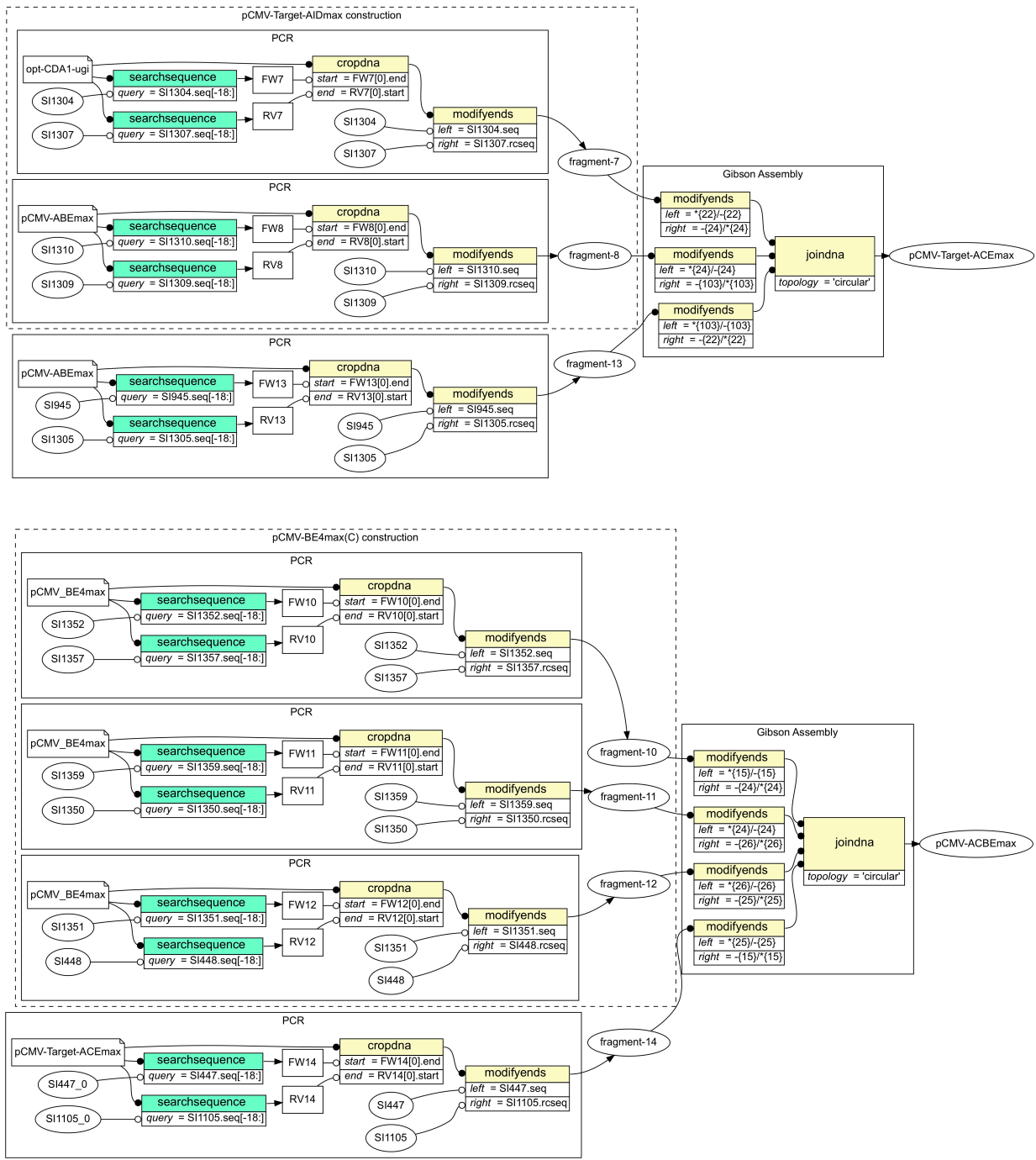

**Supplementary Figure 3**
**Operational process charts of QUEEN-generated plasmids derived by recycling previously generated protocols.** The visualizations were directly generated from the respective product gbk files by the visualizeflow function. See Supplementary Fig. 1 for detail. pCMV-Target-ACEmax and pCMV-ACBEmax were demonstrated to also be generated by recycling of protocols obtained from gbk files previously established using QUEEN. The operational histories enclosed by the dashed lines represent those recycled from the previously generated QUEEN objects.

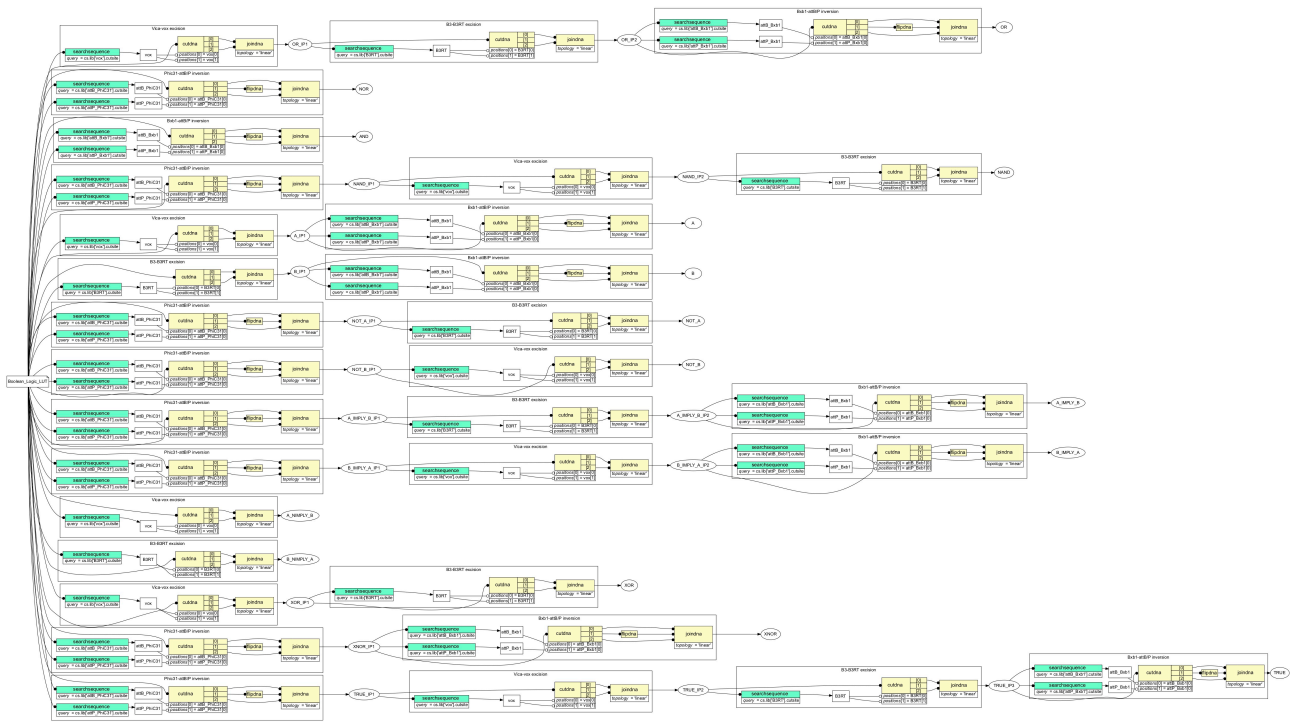

**Supplementary Figure 4**
**Operational process histories of the Boolean logic LUT circuit.** The input LUT circuit was simulated for each of the 64 input signal patterns. The visualization was directly generated from the respective product gbk file by the visualizeflow function. See Supplementary Fig. 1 for detail.

**Supplementary Table 1. List of Jupyter Notebook files**

| File name | Description | File path on the Github repository | File URL for Google Colaboratory |
|---|---|---|---|
| tutorial_ex01-23.ipynb | Example QUEEN scripts for Ex. 1 to Ex. 23. | ./demo/tutorial/tutorial_ex01-23.ipynb | https://colab.research.google.com/drive/1ubN0Q8SKXUr2t0pecu3I6Co8ctiTp0PS?usp=sharing |
| tutorial_ex24-27.ipynb | Example QUEEN scripts for Ex. 24 to Ex. 27. | ./demo/tutorial/tutorial_ex24-27.ipynb | https://colab.research.google.com/drive/1dPcNhsOl2sne_wq7ZULXXFUxizB6JQrR?usp=sharing |
| pRS112_construction.ipynb | QUEEN script for a lentiviral GFP expression plasmid construction | ./demo/sakata_et_al_2020/pRS112_construction.ipynb | https://colab.research.google.com/drive/1x3gk7gpsuAuT_ryCPxvu6CtaI7v_sIrM?usp=sharing |
| pCMV_Target_AID_construction.ipynb | QUEEN script for pCMV-Target-AID construction | ./demo/sakata_et_al_2020/pCMV_Target_AID_construction.ipynb | https://colab.research.google.com/drive/1gtgYT_Juur0DNr6atlzSBB5nnjMs_IXv_9?usp=sharing |
| pCMV_Target_ACE_construction.ipynb | QUEEN script for pCMV-Target-ACE construction | ./demo/sakata_et_al_2020/pCMV_Target_ACE_construction.ipynb | https://colab.research.google.com/drive/1JimBfiMliGn1jaYK91MtRDzzGQMjz2vf?usp=sharing |
| pCMV_Target_AIDmax_construction.ipynb | QUEEN script for pCMV-Target-AIDmax construction | ./demo/sakata_et_al_2020/pCMV_Target_AIDmax_construction.ipynb | https://colab.research.google.com/drive/1f3hiUK422a9pS0qFSkuTd4jc_IvOTktpW?usp=sharing |
| pCMV_Target_ACEmax_construction.ipynb | QUEEN script for pCMV-Target-ACEmax construction | ./demo/sakata_et_al_2020/pCMV_Target_ACEmax_construction.ipynb | https://colab.research.google.com/drive/1HwYTTnjQ-mufRpdb7Tn7NT9IpCi_-G9L?usp=sharing |
| pCMV_BE4maxC_construction.ipynb | QUEEN script for pCMV-BE4max(C) construction | ./demo/sakata_et_al_2020/pCMV_BE4maxC_construction.ipynb | https://colab.research.google.com/drive/1VIUlti8VHu2J6D65Ive5dirJp3kuFxv5?usp=sharing |
| pCMV-ACBEmax_construction.ipynb | QUEEN script for pCMV-ACBEmax construction | ./demo/sakata_et_al_2020/pCMV_ACBEmax_construction.ipynb | https://colab.research.google.com/drive/1ZR-Yq5IC9bkAf-Y4FV808iYCUJDvaivm?usp=sharing |
| pCMV_Target_ACEmax_construction_v2.ipynb | QUEEN script for pCMV-Target-ACEmax construction by editing the quine code obtained from pCMV-Target-AIDmax.gbk | ./demo/sakata_et_al_2020/pCMV_Target_ACEmax_construction_v2.ipynb | https://colab.research.google.com/drive/1MPcL4P71PpWN9_kbQoofEZtLEB0WXL9Y?usp=sharing |
| pCMV-ACBEmax_construction_v2.ipynb | QUEEN script for pCMV-ACBEmax construction by editing the quine code obtained from pCMV-BE4max(C).gbk | ./demo/sakata_et_al_2020/pCMV_ACBEmax_construction_v2.ipynb | https://colab.research.google.com/drive/1atdpNEJQPS_BiZ66SuE3nrYTvz2xfXxBw?usp=sharing |
| Boolean_logic_LUT.ipynb | QUEEN script for simulating sequence state alternations in BLADE Boolean logic LUT | ./demo/sakata_et_al_2020/Boolean_logic_LUT.ipynb | https://colab.research.google.com/drive/1sQNhWYVNXzROLQAfozoJpzQkNLHxidT8?usp=sharing |